\documentclass{article}

\usepackage{arxiv}

\usepackage[utf8]{inputenc} 
\usepackage[T1]{fontenc}    
\usepackage{hyperref}       
\usepackage{url}            
\usepackage{booktabs}       
\usepackage{amsfonts}       
\usepackage{nicefrac}       
\usepackage{microtype}      
\usepackage{lipsum}
\usepackage{graphicx}
\usepackage{amsmath}
\usepackage{textcomp}
\usepackage{gensymb}
\usepackage{siunitx}
\usepackage{cite}

\title{Capillary rise and evaporation of a liquid in a corner between a plane and a cylinder: A model of imbibition into a nanofiber mat coating}

\author{
  Noemi Ghillani\\
  Politecnico di Milano\\
  School of Industrial and Information Engineering\\
  \and
  Michael Heinz\\
  Institute for Technical Thermodynamics\\
  Technical University of Darmstadt\\
  Darmstadt, Germany\\
  \texttt{heinz@ttd.tu-darmstadt.de}\\
  \and
  Tatiana Gambaryan-Roisman\\
  Institute for Technical Thermodynamics\\
  Technical University of Darmstadt\\
  Darmstadt, Germany\\
  \texttt{gtatiana@ttd.tu-darmstadt.de}
}

\begin{document}
\maketitle

\begin{abstract}
Wetting of surfaces with porous coating is relevant for a wide variety of technical applications, such as printing technologies and heat transfer enhancement. Imbibition and evaporation of liquids on surfaces covered with porous layers are responsible for significant improvement of cooling efficiency during drop impact cooling and flow boiling on such surfaces. Up to now, no reliable model exists which is able to predict the kinetics of imbibition coupled with evaporation on surfaces with porous coatings.  In this work we consider one of possible mechanisms of imbibition on a substrate covered by a nanofiber mat. This is the capillary pressure-driven flow in a corner formed between a flat substrate and a fiber attached to it. The shape and the area of the cross-section occupied by the liquid as well as the capillary pressure change along the flow direction. A theoretical/numerical model of simultaneous imbibition and evaporation is developed, in which viscosity, surface tension and evaporation are taken into account. At the beginning of the process the imbibition length is proportional to the square root of time, in agreement with the Lucas-Washburn law. As the influence of evaporation becomes significant, the imbibition rate decreases. The model predictions are compared with experimental data for imbibition of water-ethanol mixtures into nanofiber mat coatings.
\end{abstract}

\section{Introduction}
\label{intro}

Liquid wetting of porous materials, spreading and imbibition into porous layers, as well as the concomitant heat and mass transfer phenomena play an important role in a wide variety of technical applications. Understanding complex wetting and transport phenomena is important for industrial applications such as ink-jet printing and 3D-printing, for building technologies in the control of moisture penetration into building components and filters in air conditioning systems and HVAC~\cite{BARHATE.2007}. Moreover, the knowledge of wetting behaviour and mechanisms plays a fundamental role in the development of design rules of functional materials, e.g. superhydrophobic surfaces~\cite{Deng.2012}, and of porous coatings used for cooling of electronic devices and heat transfer enhancement in spray cooling~\cite{Srikar.2009}.

When a liquid drop encounters a porous layer, the transport of liquid over and within this layer is governed the physical and chemical properties of liquid and solid in contact, the geometry of the porous structure, and the thickness of the layer. Depending on those properties, different behavior types can been observed, including perfect wetting, partial wetting, non wetting, and liquid imbibition into the layer, both in the direction parallel and normal to the substrate. The phenomenon of imbibition is present in many different types of structures and geometries that are able to cause a capillary action between liquid and solid media~\cite{Starov.2002, Bico.2001, GambaryanRoisman.2014, Alam2017imbibition, Wemp2017water, Kumar2019spreading}. The topography of the substrate, the geometry and orientation of the pores and geometrical features such as pillars can determine preferential liquid spreading directions and wetting patterns~\cite{Courbin.2009}. 

One of the forms of porous coatings, which is promising for heat transfer applications such as spray cooling and boiling, is comprised of nanofibers arranged along the substrate (nanofiber mats) \cite{Srikar.2009, Weickgenannt.2011, weickgenannt2011inverse, Wang2016water, Freystein.2016, Fischer.2017, jun2013pool, SinhaRay2017pool}. When a water drop impacts on the layer, the drop initially spreads over the surface and, as the maximum spreading diameter is reached, the drop stays pinned. The splashing and receding of drop are considerably suppressed \cite{Lembach.2010}. The  spreading stage is followed by imbibition, in which the liquid penetrates into the mat and spreads radially within the porous coating. The imbibition of water in the porous material causes the enlargement of the heat transfer area, promoting the cooling of the substrate underneath by evaporation of liquid and reducing significantly the liquid evaporation time in comparison with a drop on an uncoated substrate~\cite{Srikar.2009, Weickgenannt.2011}. The imbibed area increases, reaches a maximum and afterwards decreases due to evaporation, until the substrate is completely dried. It is clear that spreading and imbibition, from one side, and transport phenomena and phase change, from the other side, are strongly coupled.

The spontaneous liquid rise in a small cylindrical tube in contact with a reservoir can be considered as the simplest example of liquid imbibition. The liquid penetration takes place for tube diameters smaller than the capillary length $\sqrt{\sigma/\rho g}$, where $\sigma$ is the surface tension, $\rho$ is the liquid density and $g$ is the acceleration of gravity. Excluding initial stage of the capillary rise, which is governed by intertia, and the final stage, which is governed by gravity for vertical capillaries, the imbibition dynamics is dominated by the balance between viscous and capillary forces. The dynamics $l(t)$ of the capillary tubes rise is described by the Lucas-Washburn law~\cite{Washburn.1921}:
\begin{equation}
	\label{eq:washburn}
	l(t) = \sqrt{\frac{\sigma b \cos\theta}{2 \mu} t},
\end{equation}
where $\theta$ is the equilibrium contact angle, $\mu$ is the dynamic viscosity of the liquid and $b$ is the tube internal radius.

The capillary rise in porous medium or in a porous layer connected to infinite reservoir is also determined by the balance between the viscous and capillary forces, apart from the early stage of the process. Many theoretical models of imbibition are based on the assumption of Darcy flow within a porous layer or along the wall topography ~\cite{Starov.2002, Alleborn.2004, Wemp2017water}. If evaporation is not significant, the resulting imbibition length still follows the $t^{1/2}$ -  dependence, In addition, the imbibition length is proportional to the quadratic root of the permeability of the porous medium and the capillary pressure. Withstanding real system's complexity, this simple diffusive-type law appears to apply to a wide variety of porous or textured structures and rough surfaces~\cite{Bico.2001, Lembach.2010, Courbin.2009}. 

Bico et al.~\cite{Bico.2001} developed a model aiming to describe the condition of imbibition into a textured structure, formed by regularly spaced micro-size spikes. The model predicts a critical contact angle, based on the geometrical parameter of the surface design, such as surface roughness and a parameter characterizing the solid fraction remaining dry. If the condition for imbibition is satisfied, the model predicts a time evolution of the imbibition length in agreement with the $\sqrt{t}$ diffusion-like dynamics and with the results of the experiments performed on the surface by the authors.

Lembach et al.~\cite{Lembach.2010} investigated water drop impact, spreading and imbibition into a PAN electrospun nanofiber coatings. The dynamics of the radially propagating imbibition front was found once again to follow the $\sqrt{t}$-dependence, in the first stage of the process, and has been quantified by the mean transport parameter. In the experimentally observed following stages, imbibition rate decreases and eventually stops due to the combined effect of the limited liquid volume and evaporation of the fluid, which later leads to wetted area shrinkage.

In the experimental work of Courbin et al.~\cite{Courbin.2009}, drop imbibition in a texture of micropillars covering a substrate was discussed. Different flow patterns during liquid imbibition were observed, from circular to square and octagonal, underling the anisotropy of imbibition characteristic of those decorated surfaces. Furthermore, the distance of the moving contact line from the reservoir was found to generally follow the $\sqrt{t}$-dependence. Accordingly with the already observed wetting anisotropy, the moving contact line propagation prefactor $D$ in the relation $l \sim D \, t^{1/2}$ was found to assume a higher value along the diagonal of the pattern. The knowledge of micro-scale behaviours and dependencies on the geometrical parameter is, therefore, necessary for the understanding of complex wetting in texture and fiber structured substrates.

Romero and Yost~\cite{Romero.1996} and Rye et al. \cite{Rye1996flow} proposed a model for capillary-driven flow in V-shaped grooves, described by a nonlinear equation derived by mass balance and Poiseuille flow assumptions. The nonlinear equation was converted to an ordinary differential equation by similarity transformation and was later numerically solved. Results show that the wetting front position is proportional to the square root of time, and the prefactor $D$ is proportional to square root of the ratio between surface tension and viscosity of the liquid, groove depth, and a function of the contact angle and groove angle (characterizing groove opening). The authors demonstrated that the same $\sim D \, t^{1/2}$ flow dynamics can be shown in arbitrarily shaped groove. In these model evaporation of liquid was not taken into account. In addition, the influence of the meniscus shape on the relationship between the local pressure gradient and volume flow rate has been neglected.

Recently, the Computational Fluid Dynamics (CFD) methods have been applied for full-scale numerical simulations of liquid imbibition into structural elements \cite{thammanna2018computations}.  In \cite{thammanna2018computations} Volume-of-Fluid (VoF) method has been used for simulation of spontaneous propagation of a rivulet into a vertical corner. The influence of gravity has been taken into account. The results of simulations yielded the $t^{1/3}$ asymptotic for the capillary rise under gravity, which has been previously observed experimentally. This asymptotic law has been also predicted using a slender rivulet model, which allowed a similarity solution. The CFD simulations  enable the analysis of flow beyond the validity limits of the simplifying assumptions underlying approximate solutions, i.e. the inertial effect and the real meniscus shape are automatically accounted for. However, the computations are highly time consuming. No attempt has been made so far to apply full-scale numerical simulations to description of simultaneous imbibition and evaporation processes.

Only in a few works the evaporation during spreading and imbibition has been taken into account in a numerical model. Mekhitarian et al.~\cite{Mekhitarian.2017} studied experimentally and theoretically imbibition and evaporation of a volatile liquid drop into a textured surface made of cylindrical pillars (of the order of tens of micron), where the radii and heights if the pillars were varied maintaining constant distance between the pillars. The authors have found that for small pillars evaporation-dominated regime prevails, in which imbibition does not take place. In a spreading-dominated regime, taking place for large pillars, the imbibed area propagates up to a maximum extension, corresponding to the simultaneous central cap-shaped drop disappearance. The authors have developed a model describing the evolution of the drop base radius and the imbibition radius. Diffusion-limited evaporation from a thin drop has been assumed in the model. The detailed modelling of the permeability of the structure and the capillary pressure was beyond the scope of the model, and approximate expressions for these parameters have been used.   

Kolliopoulus et al. \cite{Kolliopoulos2019capillary} developed a one-dimensional model describing the imbibition and evaporation of liquids in rectangular channels. The model resulted in a scaling law for the final position of the liquid front. According to this low, the final front position is inversely proportional to the square root of evaporative mass flux.

In \cite{TGR2019imbibition} the simultaneous imbibition and evaporation from triangular grooves has been modelled for two evaporation regimes: for diffusion-governed evaporation and for evaporation driven by a constant heating rate. It was assumed that the capillary rise takes place at constant contact angle and is governed by the capillary and viscous forces. In addition, it was assumed that the imbibition length significantly exceeds the linear dimensions of the groove cross-section and that the flow in the groove is unidirectional. It was shown that, if the supply of liquid is infinite, the imbibition length first increases following the $t^{1/2}$ law and then reaches a steady state. For both evaporation regimes the maximal imbibition length is inversely proportional to the square root of the evaporation rate.

To the best of our knowledge, there have been no attempts to model the mechanisms of imbibition in porous coatings based on (nano)fibers arranged parallel to the substrate, in spite of numerous advantages of this kind of coatings for practical application. The aim of the present work is to model the liquid capillary rise coupled with evaporation within a corner between a plane (representing a substrate) and a cylindrical fiber as one of possible mechanisms contributing to imbibition on substrates coated with nanofiber mats. In order to assess the relevance of this mechanism, the results are compared with experimental data on water-ethanol mixture drop imbibition on substrates with thin nanofiber mat coatings.

\section{Physico-mathematical model}
\label{model}

\subsection{Governing equations}
\label{equations}
\begin{figure} 
	\centering 
	\includegraphics[width=.85\textwidth]{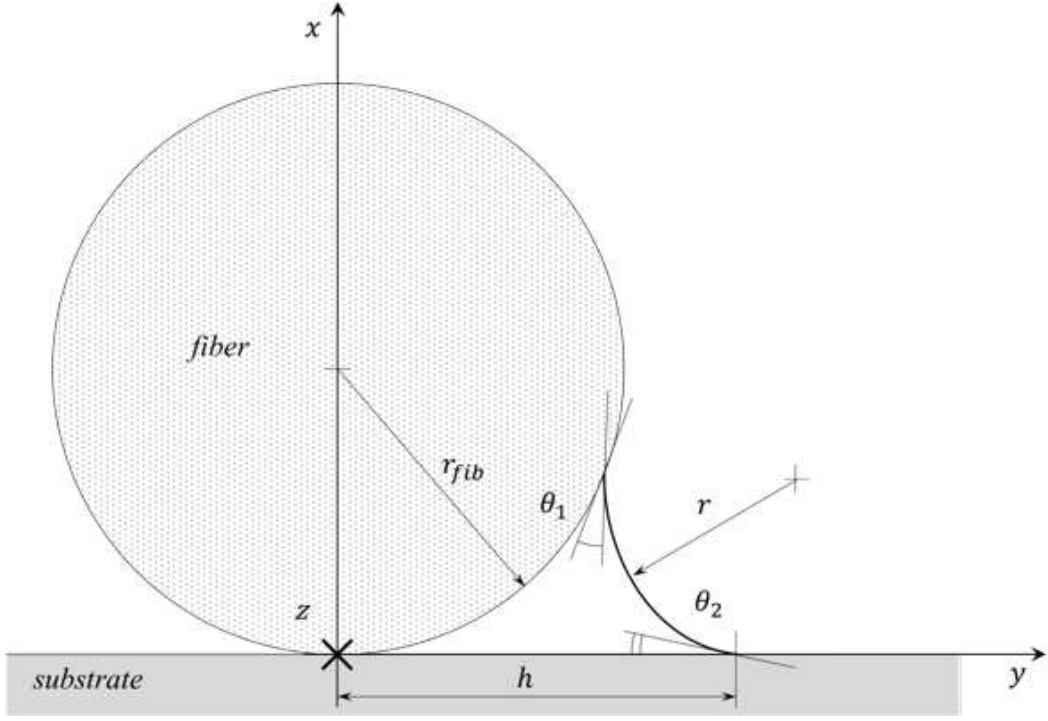}
	\caption{Schematic representation of the geometry}
	\label{fig:fib2}
\end{figure}

In this section a model describing the capillary rise in a corner between a fiber and a flat substrate coupled with liquid evaporation is introduced. In Figure~\ref{fig:fib2} a schematic representation of cross-sectional area of the modelled geometry is shown: the liquid coming into contact with the solid walls forms a meniscus whose radius $r$ depends on the contact angles $\theta_{1}$ and $\theta_{2}$, formed respectively with the fiber and the substrates.

The flow of liquid is assumed to be unidirectional over the fiber length. It is assumed that inertia and gravitational forces are negligible, and the flow is driven by the surface tension. The curvature radius $r$ is assumed to be much smaller than the meniscus curvature radius in any plane parallel to the $z$ axis, so that the interface curvature is defined in the $x-y$ plane. This curvature changes along the $z$ axis. It follows that the pressure is constant in each cross-section of rivulet and is determined by Laplace pressure jump. Thermophysical properties are assumed to be constant.

The velocity $u\left(x, y\right)$ in $z$ direction is described by the Navier-Stokes equation, which under the model assumptions accepts the following form:

\begin{equation}
\label{eqn:NSE2}
-\dfrac{\partial p}{\partial z} + \mu \left(\dfrac{\partial^2 u}{\partial x^2}+\dfrac{\partial^2 u}{\partial y^2}\right) = 0,
\end{equation}
where $p$ denotes the pressure and $\mu$ is the dynamic viscosity of the liquid. The pressure gradient can be determined from the gradient of the curvature radius:

\begin{equation}
\label{eqn:gradP2}
\dfrac{\partial p}{\partial z} = - \dfrac{\partial}{\partial z} \left( \dfrac{\sigma}{r} \right)  = \dfrac{\sigma}{r^2} \dfrac{\partial r}{\partial z},
\end{equation}
where $\sigma$ denotes the surface tension.

\begin{figure}
	\centering 
	\includegraphics[trim= 0cm 0cm 0cm 0cm, clip=true, width=.7\textwidth]{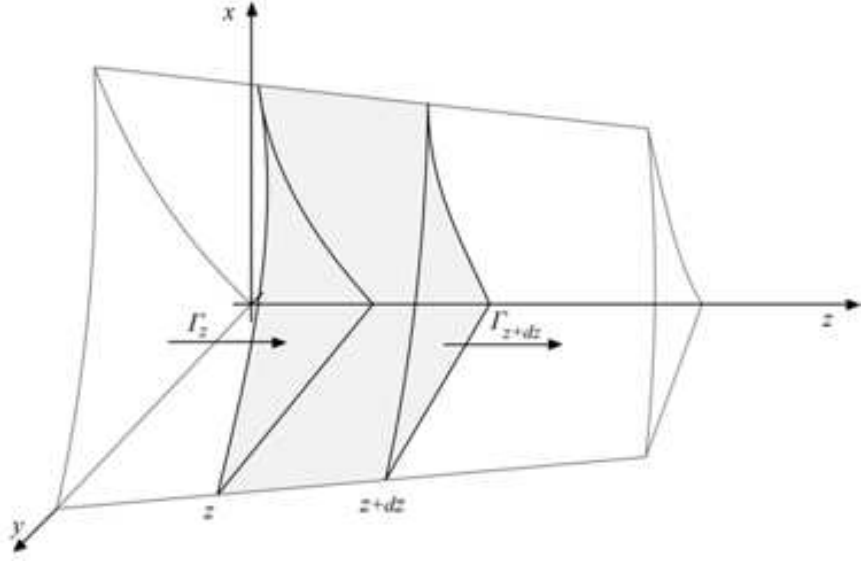}
	\caption{Control volume considered in the integral mass balance}
	\label{fig:meniscus}
\end{figure}

A part of meniscus is shown in Fig.~\ref{fig:meniscus}. The integral mass balance over the control volume $V$ illustrated in this figure has the following form:
\begin{equation}
\label{eqn:intmasbal1}
\rho \frac{\partial V}{\partial t}= \Gamma_{z} - \Gamma_{z+dz} - J_{ev}dz
\end{equation}
where $\Gamma_z$ and $\Gamma_{z+dz}$ are the mass flow rates respectively entering and exiting the control volume, $\rho$ denotes the density of the liquid and $J_{ev}$ is the evaporation rate from the given cross-section. This mass balance can be rewritten in the following form:

\begin{equation}
\label{eqn:massbalance1}
\frac{\partial a}{\partial t} = - \frac{\partial}{\partial z} \int_a u\,da - \frac{J_{ev}}{\rho},
\end{equation}
where $a$ is the cross-sectional area of the meniscus.

The velocity of the wetting front is equal to the average velocity of the liquid at this front ($z = l$):

\begin{equation}
\label{eqn:movingboundary}
\frac{d l}{d t} = \left(\frac{1}{a} \int_a u \, da \right) \, \bigg|_{z=l}.
\end{equation}

The following dimensionless variables are introduced:

\begin{eqnarray}
\label{eqn:dimensionless}
X = \frac{x}{r_{fib}}, \qquad Y = \frac{y}{r_{fib}}, \qquad Z = \frac{z}{r_{fib}}, \qquad \lambda = \frac{l}{r_{fib}}, \qquad \tau = \frac{t \sigma}{r_{fib} \mu}, \nonumber \\ H = \frac{h}{r_{fib}}, \qquad R = \frac{r}{r_{fib}}, \qquad A = \frac{a}{r_{fib}^2}, \qquad
 U = -u  \left(\frac{\sigma r_{fib}^2}{\mu r^2}  \frac{\partial r}{\partial z}\right)^{-1}.
\end{eqnarray}
where $x$, $y$ and $z$ are coordinates (see Figure~\ref{fig:fib2}), $h$ is the length of the wetted substrate (see Figure~\ref{fig:fib2}), $r$ is the meniscus radius, $u$ is the fluid velocity, and $r_{fib}$ is the radius of the fiber. We assume that the capillary rise in the corner takes place at constant contact angles $\theta_1$ and $\theta_2$. In this case the dimensionless interface curvature radius $R$ and the dimensionless cross-sectional area of the meniscus $A$ are functions of $H$.

With these definitions, the Navier-Stokes equation takes the following dimensionless form:

\begin{equation}
\label{eqn:poisson}
\frac{\partial^2 U}{\partial X^2}+\frac{\partial^2 U}{\partial Y^2} = -1.
\end{equation}

The volume flow rate can be expressed as

\begin{equation}
\label{eqn:intu}
\int_a u\,da = -\frac{\sigma r_{fib}^4}{\mu r^2}  \frac{\partial r}{\partial z} \,  \int_A U\,dA = -\frac{\sigma r_{fib}^2}{\mu R^2}  \frac{\partial R}{\partial Z} \, \, \Phi(H),
\end{equation}

where $\Phi(H) = \int_A U\,dA$ is a dimensionless flow rate.

The mass balance equation (\ref{eqn:massbalance1}) accepts the following dimensionless form:

\begin{equation}
\label{eqn:massbalance2}
\frac{\partial A}{\partial \tau} =\frac{\partial}{\partial Z} \left(\frac{\Phi}{R^2} \frac{\partial R}{\partial Z} \right) - E,
\end{equation}

or

\begin{equation}
\label{eqn:massbalance3}
\frac{d A}{d H}\frac{\partial H}{\partial \tau} = \dfrac{\Phi}{R^2} \, \dfrac{d R}{d H} \, \dfrac{\partial^2 H}{\partial Z^2} + \dfrac{d}{d H} \biggl(\dfrac{\Phi}{R^2} \, \dfrac{d R}{d H}\biggr) \, \biggl(\dfrac{\partial H}{\partial Z}\biggr)^2 - E,
\end{equation}

where
\begin{equation}
\label{eqn:E}
E = \frac{J_{ev} \nu}{r_{fib} \sigma} 
\end{equation}
is the dimensionless evaporation number.

The equation for the velocity of the wetting front propagation (\ref{eqn:movingboundary}) has the following dimensionless form:

\begin{equation}
\label{eqn:movingboundary2}
\frac{d \lambda}{d \tau} = -\left(\frac{\Phi}{AR^2} \frac{\partial R}{\partial Z} \right) \, \bigg|_{Z=\lambda} = -\left(\frac{\Phi}{AR^2} \frac{d R}{d H}\frac{\partial H}{\partial Z} \right) \, \bigg|_{Z=\lambda}.
\end{equation}

Equation (\ref{eqn:massbalance2}) is defined in the interval $0\leq Z \leq \lambda(\tau)$, which changes with the time. Introducing the new dimensionless coordinate

\begin{equation}
\label{eqn:psi}
\psi = \frac{Z}{\lambda(\tau)} 
\end{equation}
and performing the transformation of variables, the mass balance equation can be brought to the form:

\begin{equation}
\label{eqn:massbalance4}
\frac{d A}{d H}\biggl(\lambda^2\frac{\partial H}{\partial \tau} - \frac{1}{2}\frac{d\lambda^2}{d\tau}\psi \dfrac{\partial H}{\partial \psi} \biggr) = \dfrac{\Phi}{R^2} \, \dfrac{d R}{d H} \, \dfrac{\partial^2 H}{\partial \psi^2} + \dfrac{d}{d H} \biggl(\dfrac{\Phi}{R^2} \, \dfrac{d R}{d H}\biggr) \, \biggl(\dfrac{\partial H}{\partial \psi}\biggr)^2 - \lambda^2E.
\end{equation}

This equation is defined in the interval $0 \leq \psi \leq 1$. The equation for the velocity of the wetting front propagation has the form:

\begin{equation}
\label{eqn:movingboundary3}
\frac{1}{2}\frac{d \lambda^2}{d \tau} = -\left(\frac{\Phi}{AR^2} \frac{d R}{d H}\frac{\partial H}{\partial \psi} \right) \, \bigg|_{\psi=1}.
\end{equation}

By introducing the functions

\begin{eqnarray}
\label{eqn:coefficients}
c_{1}(H)= \frac{\Phi}{R^2} \frac{d R}{d H} \biggl(\frac{d A}{ d H}\biggr)^{-1},   \nonumber \\
c_{2}(H)= \frac{d}{d H} \biggl(\frac{\Phi}{R^2} \frac{d R}{d H}\biggr) \biggl(\frac{d A}{ d H}\biggr)^{-1},   \nonumber \\
c_{3}(H)= \frac{\Phi}{A\,R^2} \, \frac{d R}{d H},
\end{eqnarray}
Eqs. (\ref{eqn:massbalance4}, \ref{eqn:movingboundary3}) can be written down in a compact form:

\begin{equation}
\label{eqn:massbalance5}
\frac{\partial H}{\partial \tau} = \frac{\psi}{2\lambda^2}\frac{d\lambda^2}{d\tau} \dfrac{\partial H}{\partial \psi}  + \dfrac{c_1}{\lambda^2} \, \dfrac{\partial^2 H}{\partial \psi^2} + \dfrac{c_2}{\lambda^2} \biggl(\dfrac{\partial H}{\partial \psi}\biggr)^2 - E \biggl(\frac{d A}{ d H}\biggr)^{-1},
\end{equation}

\begin{equation}
\label{eqn:movingboundary4}
\frac{1}{2}\frac{d \lambda^2}{d \tau} = -c_3(0)\left(\frac{\partial H}{\partial \psi} \right) \, \bigg|_{\psi=1}.
\end{equation}

The functions $H(\psi, \tau)$ and $\lambda(\tau)$ can be determined by solution of the system of equations (\ref{eqn:massbalance4}, \ref{eqn:movingboundary3}) subject to the boundary conditions

\begin{equation}
\label{eqn:bc}
H(0, \tau) = H_m, \qquad H(1, \tau) = 0
\end{equation}

and initial conditions

\begin{equation}
\label{ic}
H(\psi, 0) = H_m (1 - \psi), \qquad \lambda(0) = \lambda_{min}.
\end{equation}

If in the above $\lambda_{min}$ is a small enough, the choice of $\lambda_{min}$ and of the initial distribution of $H$ do not affect the long-time evolution of the functions $\lambda(\tau)$ and $H(\psi, \tau)$. $H_m$ is maximum value of the dimensionless wetted substrate length $H$, which corresponds to infinite curvature radius for the fixed values of $\theta_1$ and $\theta_2$:

\begin{equation}
\label{Hm}
H_m = 2 \frac{\cos{\frac{\theta_2-\theta_1}{2}}\cos{\frac{\theta_2+\theta_1}{2}}}{\sin{\theta_2}}.
\end{equation}

By setting the boundary condition $H(0, \tau) = H_m$ we assume that the liquid is supplied to the corner between the substrate and the fiber from an infinite reservoir, for example, a large drop.

\subsection{Solution}
\label{solution}

\begin{figure}
	\centering 
	\includegraphics[trim= 0cm 0cm 0cm 0cm, clip=true, width=.7\textwidth]{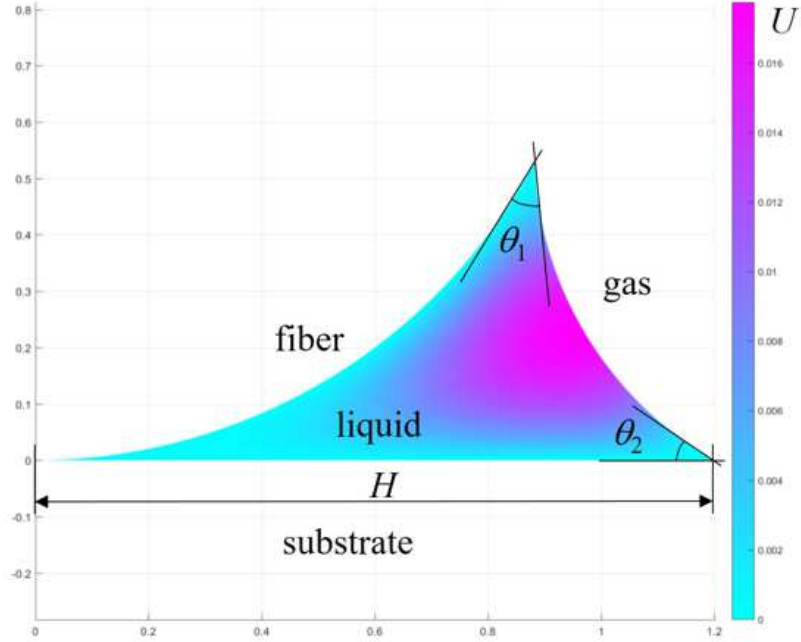}
	\caption{Solution of Poisson equation in a meniscus cross-section for $\theta_1 = \theta_2 = \pi/6$, $H=1.2$.}
	\label{fig:PoissonSol}
\end{figure}

The velocity field in a meniscus cross-section has been determined by the solution of Eq. (\ref{eqn:poisson}). The geometry of the computational domain is fully determined by the contact angles $\theta_1$ and $\theta_2$ and by the dimensionless wetted substrate length $H$. The latter varies between zero and $H_m$. An example of the computational domain together with the resulting solution for the dimensionless velocity field $U(X, Y)$ is illustrated in Fig. \ref{fig:PoissonSol}. The Poisson equation has been solved subject to the no-slip boundary condition (homogeneous Dirichlet conditions) at the substrate/liquid and fiber/liquid interfaces and zero shear stress boundary condition (homogeneous Neumann condition) at the liquid/gas interface.

The Poisson equation (\ref{eqn:poisson}) has been solved using the Finite Elements Method (FEM) within the Matlab PDE Toolbox. The resulting field has been numerically integrated over the liquid area $A$  to obtain $\Phi(H)$. The convergence test has been performed by mesh refinement. The number of elements has been chosen in a such a way that the relative change of value of $\Phi$ after the further mesh refinement is below $10^{-5}$.

The functions $R(H)$ and $A(H)$ have been determined by the numerical solution of a system of relevant trigonometric equations. On the basis of known $R(H)$, $A(H)$ and $\Phi(H)$ the auxiliary functions $(dA/dH)^{-1}$, $c_1(H)$, $c_2(H)$ and $c_3(H)$ have been evaluated following the definitions (\ref{eqn:coefficients}). These functions show the following asymptotic behavior as $H$ approaches zero:

\begin{eqnarray}
\label{eqn:asymptotic}
\biggl(\frac{d A}{ d H}\biggr)^{-1} \propto H^{-2},   \nonumber \\
c_{1}(H) \propto H^2,   \nonumber \\
c_{2}(H) \propto H,   \nonumber \\
c_{3}(H) \propto H.
\end{eqnarray}

It is evident that $\psi=1$ is a singular point of Eqs. (\ref{eqn:massbalance5}, \ref{eqn:movingboundary4}). In particular, it follows from Eq. (\ref{eqn:movingboundary4}) that, as soon as $d\lambda^2/d\tau \neq 0$, the derivative $\frac{\partial H}{\partial \psi}$ is infinite at $\psi=1$. In order to handle this singularity, we investigate the behavior of the solution in the vicinity of the point $\psi=1$. Equation (\ref{eqn:movingboundary4}) can be rewritten in the form:

\begin{equation}
\label{eqn:gamma}
\frac{1}{2}\frac{d \lambda^2}{d \tau} = -\gamma\left(H\frac{\partial H}{\partial \psi} \right) \, \bigg|_{\psi=1} = -\frac{\gamma}{2}\left(\frac{\partial H^2}{\partial \psi} \right) \, \bigg|_{\psi=1} ,
\end{equation}
where the coefficient $\gamma$ depends on the values of $\theta_1$ and $\theta_2$. This leads to the following asymptotic behavior in the vicinity of the point $\psi=1$:

\begin{eqnarray}
\label{eqn:Hbehavior}
\left(\frac{\partial H^2}{\partial \psi} \right) \, \bigg|_{\psi \to 1} \to \frac{1}{\gamma}\frac{d \lambda^2}{d \tau},  \nonumber \\
H^2 \, \bigg|_{\psi \to 1} \to -\frac{1}{\gamma}\frac{d \lambda^2}{d \tau} \left( 1 - \psi\right) . 
\end{eqnarray}

 Combining these two equations, we get
 
 \begin{equation}
\label{eqn:inbetween}
\left(\frac{1}{H^2}\frac{\partial H^2}{\partial \psi} \right) \, \bigg|_{\psi \to 1} \to \left(\frac{2}{H}\frac{\partial H}{\partial \psi} \right) \, \bigg|_{\psi \to 1} \to -\frac{1}{1-\psi}.
\end{equation}
 
 Equation (\ref{eqn:inbetween}) can be used for definition of a boundary condition of the third kind at $\psi=1-\varepsilon$, where $\varepsilon<<1$:
 
 \begin{equation}
\label{eqn:bcepsilon}
\frac{\partial H}{\partial \psi} + \frac{H}{2 \varepsilon} = 0
\end{equation}

The system of equations (\ref{eqn:massbalance5}, \ref{eqn:movingboundary4}) has been solved in the domain $0 \leq \psi \leq 1 - \varepsilon$ subject to the first boundary condition of Eq. (\ref{eqn:bc}) and the boundary condition (\ref{eqn:bcepsilon}).

\section{Experimental method}
\label{experiment}

\begin{figure}
	\centering 
	\includegraphics[trim= 0cm 0cm 0cm 0cm, clip=true, width=.7\textwidth]{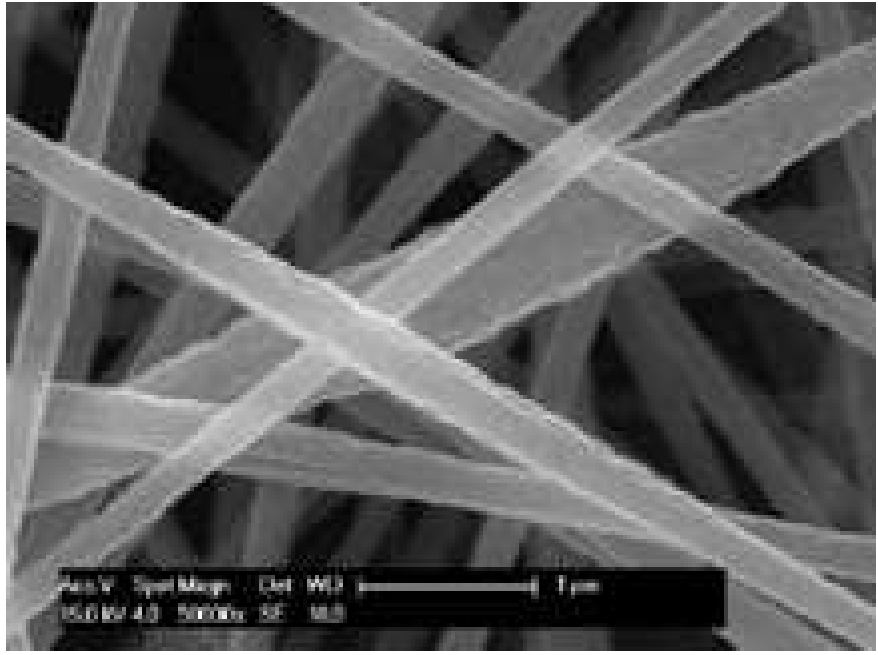}
	\caption{SEM image of nanofiber mats.}
	\label{fig:nanofibers}
\end{figure}

Nanofiber coatings have been manufactured by electrospinning. The electrospinning setup used in this work was purchased from Avectas Spraybase. In this work square silicon wafers  have been used as base substrates. Polyacrylonitrile (PAN), a polymer that is partially wettable by water \cite{Lembach.2010}, has been used to spin the nanofiber mats. The spinning solution consists of 5 wt.\% PAN dissolved in DMF (Dimethylformamide). The solution was stirred for 48 hours at room temperature and set to rest for an additional two hours prior to electrospinning. Fiber diameters for the 5 wt.$\%$ PAN/DMF solution are of the order of 200-300 nanometers with some bead-like structures stretched over several micrometers. The pore sizes at the surface are of the order of $4-7$ $\mathrm{\mu m}$. The mat thickness used in this work was around $45-50$ $\mathrm{\mu m}$. A typical SEM image of the nanofiber mat is shown in Fig. \ref{fig:nanofibers}.

Drop spreading experiments have been carried out in a closed test cell in air atmosphere at a temperature of 26 \textdegree C and a relative humidity of $\approx 2.5\,\%$. Drops are dispensed with a syringe pump which is connected to a canula inside the test cell. The coated substrates are placed underneath the canula. The distance of the needle tip to the substrate surface is $10\,\mathrm{mm}$. The experiments have been performed for pure water, pure ethanol and their mixtures for 10, 20, 50, 80 and 90 wt.$\%$ of ethanol. The dynamic viscosity of the mixtures have been measured with the Brookfield DV-III ULTRA instrument. The volume of the dispensed drops, is determined by the balance between gravity and surface tension, depended on composition. The initial drop diameter was 2.43 mm for pure water, 1.81 mm for 50 wt.$\%$ of ethanol, and 1.75 mm for pure ethanol. The drop spreading and evaporation experiments have been repeated nine times for each mixture. The imbibition into the nanofiber mat and consequent evaporation have been observed with a camera (Andor Zyla 5.5)  mounted in a top view with a frame rate of 36 Hz.

\section{Results and discussion}
\label{results}

\subsection{Numerical results}
\label{num_res}

\begin{figure}
	\centering 
	\includegraphics[trim= 0cm 0cm 0cm 0cm, clip=true, width=.9\textwidth]{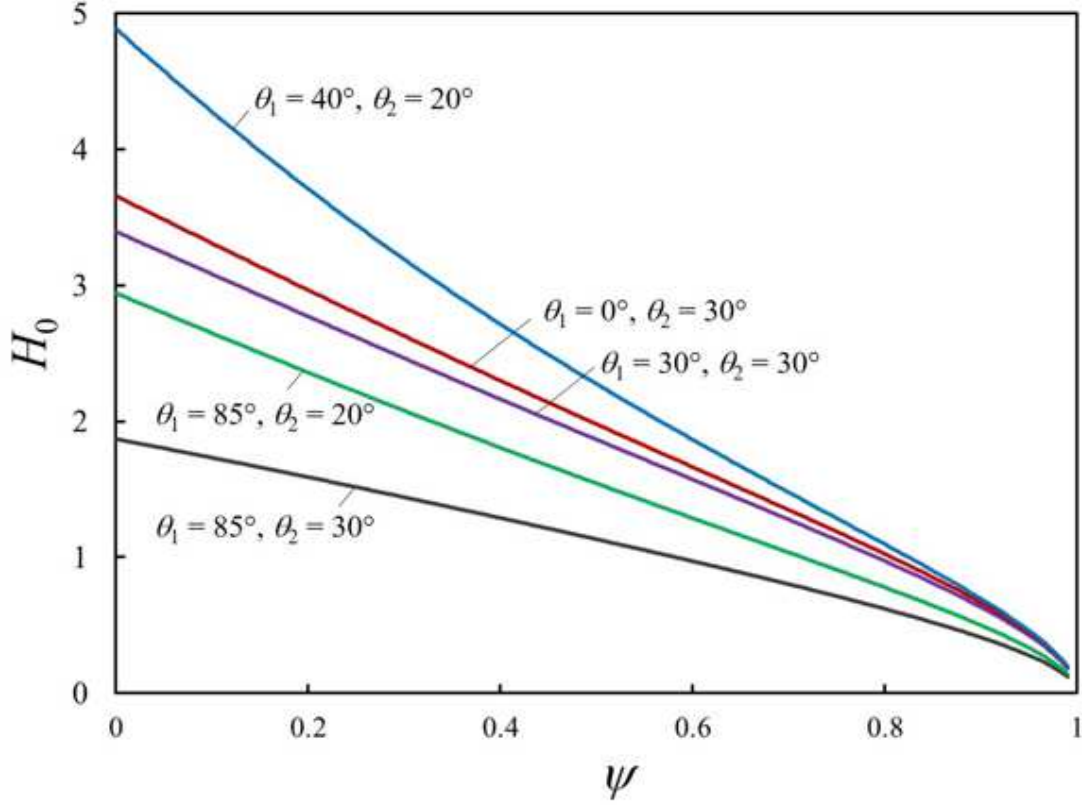}
	\caption{Time-independent distributions of dimensionless wetted length, $H_0$, for different values of contact angles $\theta_1$ and $\theta_2$ (results of numerical simulation).}
	\label{fig:profiles}
\end{figure}

\begin{figure}
	\centering 
	\includegraphics[trim= 0cm 0cm 0cm 0cm, clip=true, width=.9\textwidth]{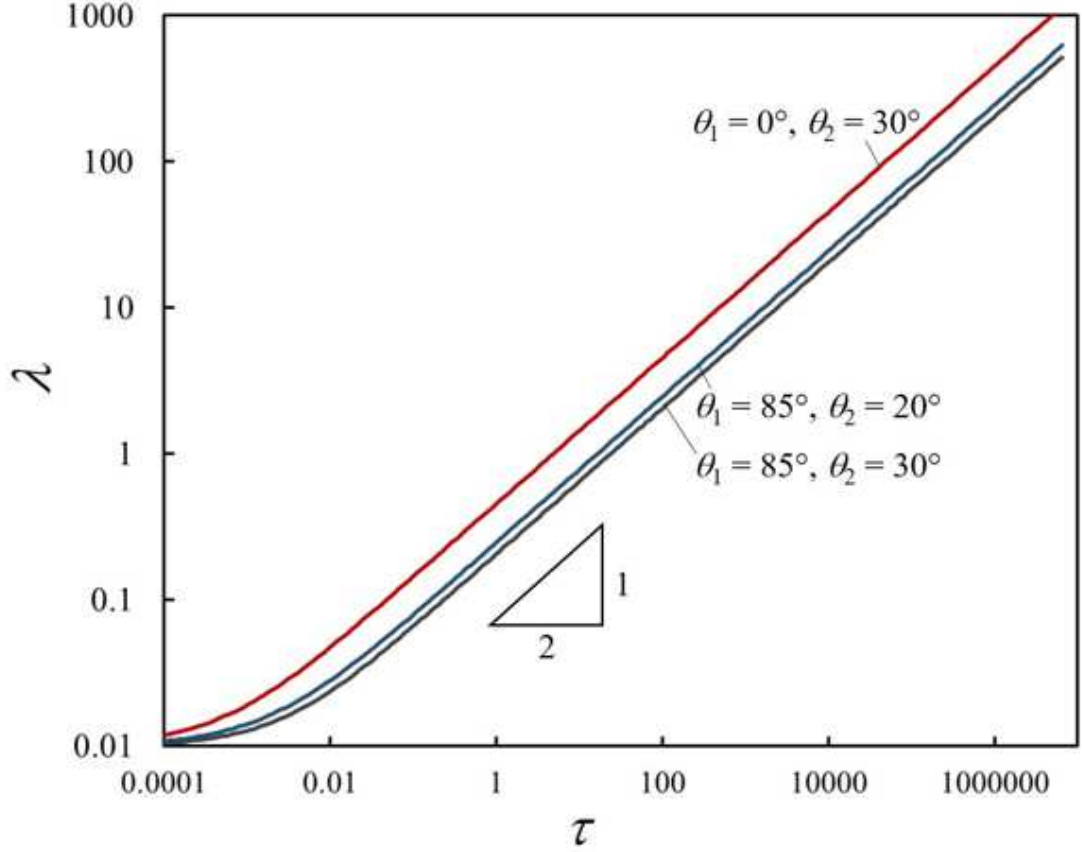}
	\caption{Dimensionless imbibition length for different values of contact angles $\theta_1$ and $\theta_2$ in the absence of evaporation (results of numerical simulation).}
	\label{fig:lambda_no_evap}
\end{figure}

The solution of Eqs. (\ref{eqn:massbalance5}, \ref{eqn:movingboundary4}) in the absence of evaporation ($E=0$) tends asymptotically to a time-independent self-similar distribution $H_0\left(\psi\right)$. This distribution is depicted in Fig.~\ref{fig:profiles} for different values of contact angles $\theta_1$ and $\theta_2$. The temporal evolution of dimensionless imbibition front position for these contact angles is shown in Fig. \ref{fig:lambda_no_evap}. The plots corresponding to the sets $(\theta_1=0 \degree, \theta_2=30\degree )$, $(\theta_1=40\degree , \theta_2=20\degree )$  and $(\theta_1=30\degree, \theta_2=30\degree)$ are indistinguishable from each other on this scale. It is obvious that all the $\lambda(\tau)$ curves asymptotically tend to straight lines in logarithmic scale with $\lambda \propto \tau^{1/2}$, indicating the Lucas-Washburn imbibition law \cite{Washburn.1921}. Similar trends have been predicted for the capillary rise in a triangular groove \cite{Rye1996flow, TGR2019imbibition}. The asymptotic behavior of the $\lambda(\tau)$ dependence is reached as soon as the the distribution $H\left(\psi, \tau\right)$ attains the time-independent form $H_0\left(\psi\right)$. Moreover, the coefficient of proportionality in the asymptotic relation $\lambda \propto \tau^{1/2}$ can be determined from the following asymptotic form of equations (\ref{eqn:massbalance5}, \ref{eqn:movingboundary4}) with $E=0$:

\begin{equation}
\label{eqn:massbalance6}
c_1 \, \dfrac{d^2 H_0}{d\psi^2} + c_2 \biggl(\dfrac{d H_0}{d\psi}\biggr)^2 = -\psi\frac{ \kappa^2}{2} \dfrac{d H_0}{d \psi} ,
\end{equation}

\begin{equation}
\label{eqn:movingboundary5}
\frac{\kappa^2}{2} = -c_3(0)\left(\frac{d H_0}{d \psi} \right) \, \bigg|_{\psi=1}.
\end{equation}

The parameter 

\begin{equation}
\label{eqn:kappa}
\kappa = \left(\lim\limits_{\tau \to \infty}\frac{d\lambda^2}{d\tau}\right)^{1/2}
\end{equation}

determines the rate of capillary rise in the Lucas-Washburn regime. The dimensional capillary rise dynamics in this regime can be determined from the relation

\begin{equation}
\label{eqn:lWashburn}
l^2(t) = l_0^2 + \kappa^2 \frac{t r_{fib} \sigma}{\mu}.
\end{equation}

\begin{figure}
	\centering 
	\includegraphics[trim= 0cm 0cm 0cm 0cm, clip=true, width=.8\textwidth]{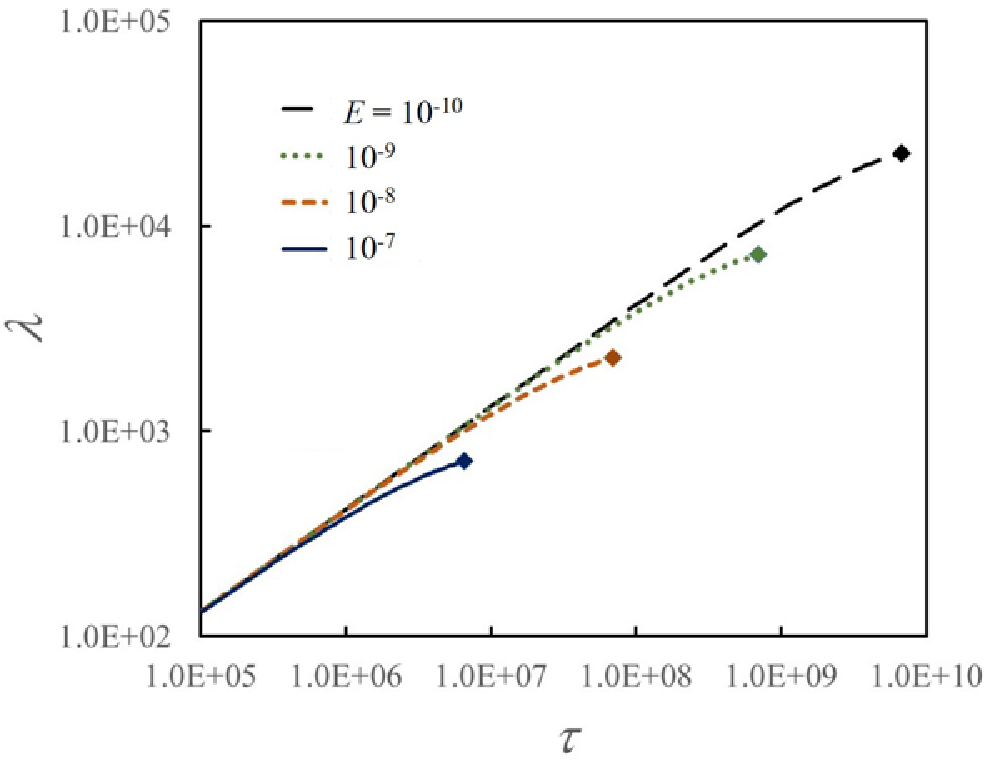}
	\caption{Dimensionless imbibition length for $\theta_1 = 30 \degree$ and $\theta_2= 30 \degree$ and different values of $E$ (results of numerical simulation). The diamond-shaped symbols correspond to the points ($\tau_{max}, \lambda_{max}$).}
	\label{fig:lambda_evap}
\end{figure}

\begin{figure}
	\centering 
	\includegraphics[trim= 0cm 0cm 0cm 0cm, clip=true, width=1.0\textwidth]{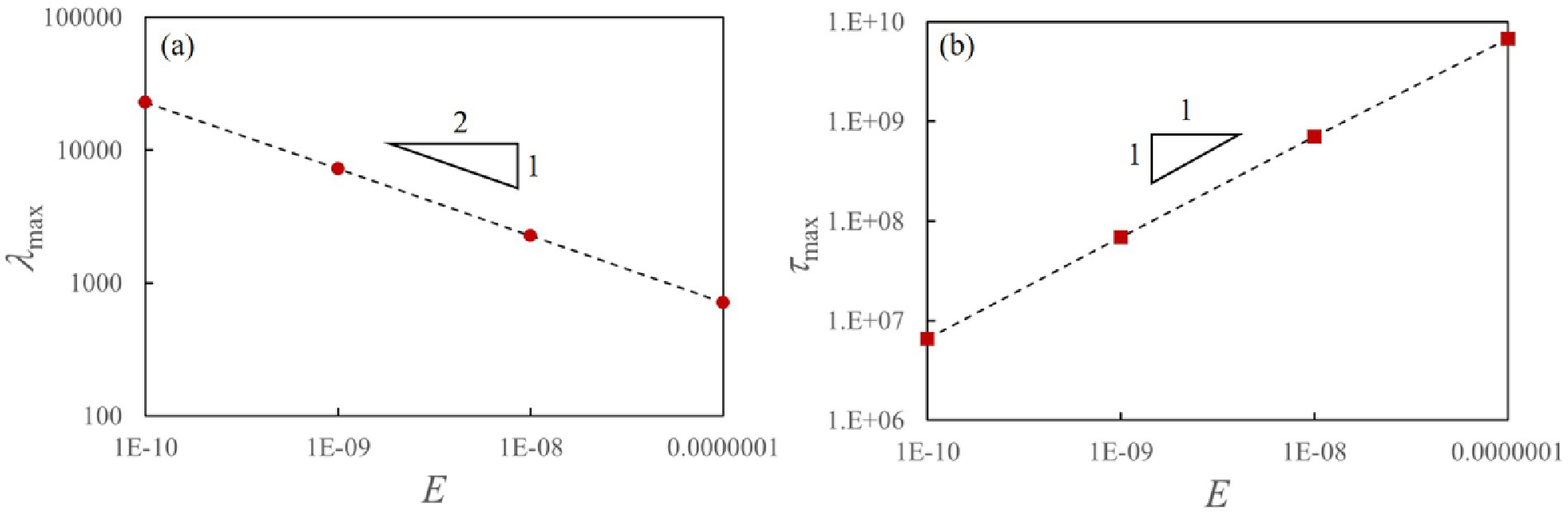}
	\caption{Nominal maximal dimensionless imbibition length (a) and dimensionless time, at which this length is reached (b) as functions of dimensionless evaporation rate, $E$.}
	\label{fig:evap_summary}
\end{figure}

In the presence of evaporation the rate of capillary rise decreases due to the mass loss. The evolution of dimensionless imbibition length for $\theta_1 = 30 \degree$ and $\theta_2= 30 \degree$ is depicted in Fig. \ref{fig:lambda_evap} for different values of the dimensionless evaporation rate, $E$. It is evident that, starting from a certain time instant, the evolution $\lambda(\tau)$ deviates from the Lucas-Washburn dynamics. With increasing of $E$ the deviation from the Lucas-Washburn regime takes place earlier. In the case of simultaneous capillary rise and evaporation in triangular grooves it has been shown that at large values of $\tau$ the rivulet length reaches a maximum value, which is inversely proportional to the square root of dimensionless evaporation rate \cite{TGR2019imbibition}. This result has been obtained both for evaporation at constant wall heat flux leading to constant evaporation rate, and for the diffusion-limited evaporation. In the present work we have defined the nominal maximal dimensionless imbibition length based on the condition that the rate of change of the imbibition length abruptly decreases. We have found that the most robust criterion for that is the time derivative $\frac{d\lambda^2}{d\tau}$. Therefore, the nominal maximal imbibition length, $\lambda_{max}$, and the time at which this length is reached, $\tau_{max}$, are defined at the instant for which $\frac{d\lambda^2}{d\tau}=0.03$. These points are marked in Fig. \ref{fig:lambda_evap} by diamond symbols for all $\lambda(\tau)$ curves. The values $\lambda_{max}$ and $\tau_{max}$ are depicted in Fig.~\ref{fig:evap_summary} as functions of $E$. It is clear that $\lambda_{max} \propto E^{-1/2}$ and $\tau_{max} \propto E$. The same scaling has been predicted for the capillary rise in triangular grooves in \cite{TGR2019imbibition} and for capillary rise and evaporation in open rectangular microchannels \cite{Kolliopoulos2019capillary}. Evidently, the results obtained in this section are only valid for an infinite supply of liquid at the origin ($z=\psi=0$).

\subsection{Experimental results and comparison with the model}
\label{exp_res}

\begin{figure}
	\centering 
	\includegraphics[trim= 0cm 0cm 0cm 0cm, clip=true, width=.9\textwidth]{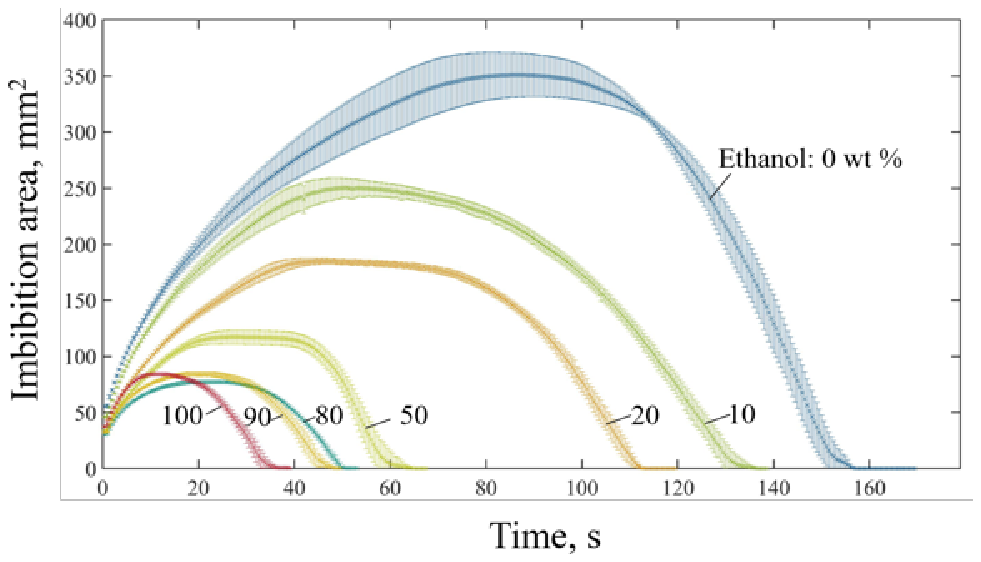}
	\caption{Measured imbibed area as a function of time for water-ethanol mixtures.}
	\label{fig:exp_data}
\end{figure}

\begin{figure}
	\centering 
	\includegraphics[trim= 0cm 0cm 0cm 0cm, clip=true, width=.9\textwidth]{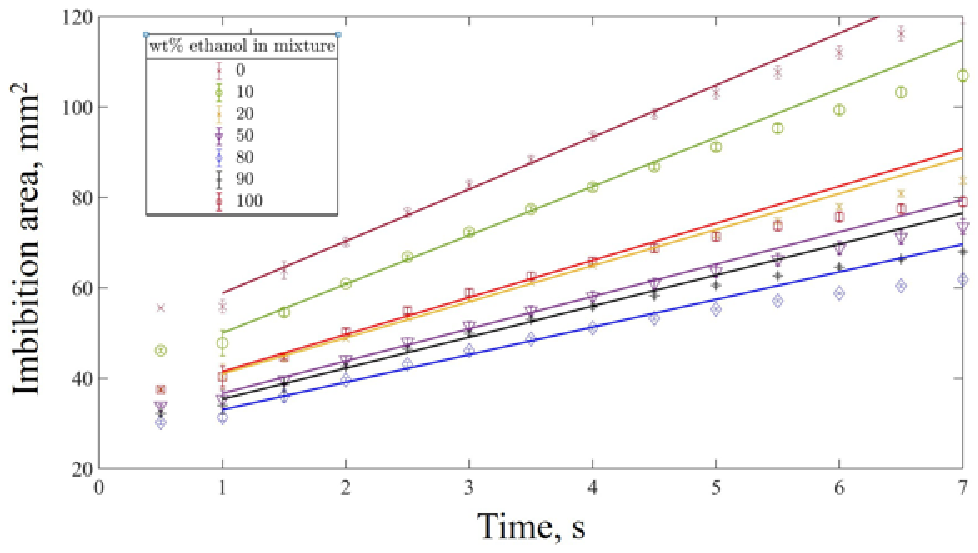}
	\caption{Measured imbibed area as a function of time for water-ethanol mixtures for the early stage of imbibition. The straight lines are based on a linear fitting of the data in the interval from 1s to 5s for all compositions.}
	\label{fig:exp_early}
\end{figure}

The measured imbibition area as a function of time is presented in Fig. \ref{fig:exp_data}. For all drop compositions the same scenario can be observed: the imbibition area initially increases, reaches a maximum value, decreases due to evaporation and finally reaches zero. Since in the case of a drop spreading the supply of liquid is limited by the drop volume, the maximal imbibition area can be determined by evaporation rate or by the drop volume. Each curve (surrounded by error bars) corresponding to a certain composition of liquid results from a series of nine test runs. The relatively wide error bars for the imbibition curves with water and the water-ethanol mixtures with high water concentration appear due to a deformation of the nanofiber mats by the capillary forces. The maximal imbibition area decreases for each subsequent spreading drop. However, this scattering of data is small for the times below 10s. The measured imbibition area at the early stage of experimental runs (without the error bars) is shown in Fig.~\ref{fig:exp_early} for the early stage of imbibition. The straight lines show the linear fitting in the interval between 1s to 5s. It is clear that within this time interval the imbibition follows the Lucas-Washburn law:

\begin{equation}
\label{eqn:lexpWashburn}
a_{imb}(t) = \pi l_{exp}^2(t) = a_{imb,0} + kt,
\end{equation}
where $a_{imb}$ denotes the imbibition area, $l_{exp}$ is the equivalent imbibition length, $a_{imb,0}$ and $k$ are the parameters of the linear fit. The linear dependence between the imbibition area and the time implies that in this time interval evaporation does not affect the imbibition rate. Based on this fitting and comparison with Eq. (\ref{eqn:lWashburn}), one can define the parameter

\begin{equation}
\label{eqn:expkappa}
\kappa_{exp} = \frac{k\mu}{\pi r_{fib}\sigma}.
\end{equation}

This value can be used for assessment of the governing imbibition mechanisms. In the following we use the value $\kappa_{exp}$ in order to assess the relevance of the capillary rise of the liquid between the substrate and the first layer of the nanofibers to the imbibition of liquid into the nanofiber mat coatings. It is clear that the imbibition can depend on multiple additional mechanisms, including the capillary rise in "channels" formed by nanofibers of different layers which are not in contact with the silicon substrate. In addition, the fibers in the coating are oriented randomly and not only along the path from the drop center to the periphery. However, we expect that, if the mechanism described in the theoretical part of this work has a significant contribution to the overall imbibition process, the $\kappa_{exp}$ values are close or at least comparable with the numerically predicted values of $\kappa$.

The properties of the water-ethanol mixture, the values of $k$ determined from the linear fit to experimental data presented in Fig. \ref{fig:exp_early} and the values of $\kappa_{exp}$ determined from Eq. (\ref{eqn:expkappa}) with $r_{fib}=400$ nm are summarized in Table \ref{tab:1}. The dynamic viscosity has been measured by the authors, and the surface tension data at 25 $\degree$C have been obtained from the literature \cite{Vazquez1995surface}. The values of the contact angles between the water-ethanol mixture and the silicon substrate (corresponding to $\theta_2$) have been interpolated from \cite{Spencer2013contact} for the case of untreated air-aged silicon. No data on the contact angle between the ethanol-water mixture and PAN nanofibers could be found in literature. It is known however that the surface of the nanofibers is rough (see Fig. \ref{fig:nanofibers}) and that the water on PAN cast samples exhibits the contact angle of 30-40 $\degree$ \cite{Lembach.2010}. Since the surface tension of ethanol is significantly below the surface tension of water, we can suggest that the angle $\theta_1$ for all water-ethanol mixtures lies below 40 $\degree$.

\begin{table}
\caption{Mixture properties and imbibition parameters.}
\label{tab:1}       
\begin{tabular}{llllll}
\hline\noalign{\smallskip}
Ethanol wt.\% & $k$, mm$^2$/s & $\mu$, kg/(m s) & $\sigma$, kg/s$^2$ & $\kappa_{exp}$  & $\theta_2,$ \degree \\
\noalign{\smallskip}\hline\noalign{\smallskip}
0 & 11.54 & $8.83\cdot10^{-4}$  & $72.01\cdot10^{-3}$ & 0.336 &  47.7 \\
10 & 10.77 & $1.24\cdot10^{-3}$  & $47.53\cdot10^{-3}$ & 0.473 &  41.6 \\
20 & 7.97 & $1.70\cdot10^{-3}$  & $37.97\cdot10^{-3}$ & 0.532 &  36.9 \\
50 & 7.14 & $2.17\cdot10^{-3}$  & $27.96\cdot10^{-3}$ & 0.664 &  23.9 \\
80 & 6.10 & $1.91\cdot10^{-3}$  & $23.82\cdot10^{-3}$ & 0.624 &  16.6 \\
90 & 6.86 & $1.53\cdot10^{-3}$  & $22.72\cdot10^{-3}$ & 0.606 &  15.3 \\
100 & 8.20 & $1.08\cdot10^{-3}$  & $21.82\cdot10^{-3}$ & 0.568 &  14.8 \\
\noalign{\smallskip}\hline
\end{tabular}
\end{table}

It can be observed in Table \ref{tab:1} that the the values $k$ and $\kappa_{exp}$ show different trends. The parameter $k$ exhibits a minimum for the mixture with 80 $\%$ ethanol mass fraction, whereas $\kappa_{exp}$ has a maximum at 50 $\%$ ethanol mass fraction. This is explained by the nonmonotonic behaviour of dynamic viscosity of the water-ethanol mixture, which has a strong maximum at 50 $\%$ ethanol mass fraction.

\begin{figure}
	\centering 
	\includegraphics[trim= 0cm 0cm 0cm 0cm, clip=true, width=.9\textwidth]{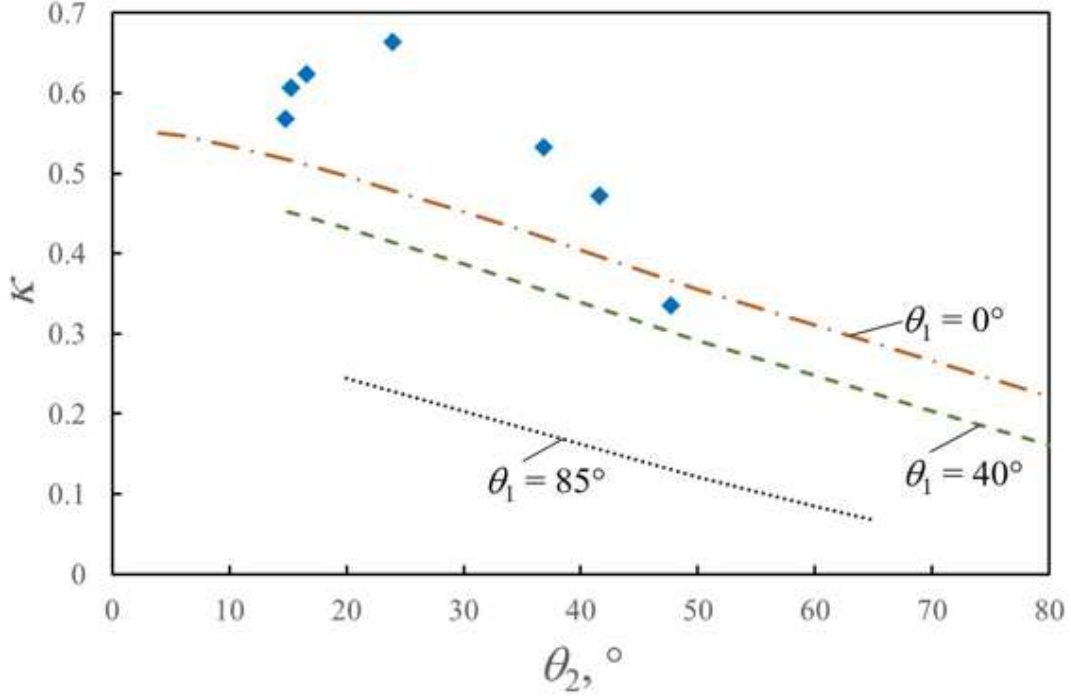}
	\caption{Imbibition front propagation coefficient as a function of the contact angle $\theta_2$. Lines represent the results of numerical simulations, diamond-shaped symbols illustrate the experimental data (see the column $\kappa_{exp}$ in Table \ref{tab:1}).}
	\label{fig:Comparison}
\end{figure}

The experimentally determined values $\kappa_{exp}$ along with the results of the simulations are displayed in Fig. \ref{fig:Comparison} versus the contact angle $\theta_2$. The results of the simulations are shown for $\theta_1=85 \degree$, $\theta_1=40 \degree$ and $\theta_1=0 \degree$ (perfect wetting). Evidently, the predicted values of $\kappa$ decrease monotonically with increasing of $\theta_1$ and $\theta_2$. The experimental data are shown using the diamond-shaped symbols. First of all, it can be seen that the experimental data are close to the model predictions for $\theta_1=0 \degree$. This confirms that the capillary rise in a corner between the substrate plane and the nanofibers from the first layer is a mechanism which possibly contributes to the imbibition into the nanofiber mat coating. However, the maximum of $\kappa_{exp}$ for the mixture with 50 wt $\%$ ethanol can't be explained in the framework of the suggested mechanism. One of possible explanations of nonmonotonic behavior of $\kappa_{exp}$ is the effect of evaporation combined with Marangoni effect. 

\section{Conclusions}
\label{Conclusions}

In this work a theoretical model is developed which describes simultaneous capillary rise and evaporation in a corner formed between a plane and a cylinder. This is one of the mechanisms contributing to imbibition of liquids into nanofiber mat coatings. It is assumed that the flow in the corner is unidirectional, the imbibition length is much larger than the fiber radius and that the flow is governed by the surface tension and viscosity. In addition, the constant and uniform evaporation rate is assumed. The model predicts that, as long as the evaporation is insignificant, the imbibition length increases proportional to the square root of time, in agreement with the Lucas-Washburn law. Evaporation leads to decreasing of the capillary rise rate. The predicted nominal maximal dimensionless imbibition length is inversely proportional to the square root of the dimensionless evaporation rate, which agrees with the scaling predicted for capillary rise in triangular grooves and for capillary rise in open rectangular microchannels.

Imbibition of ethanol, water and water-ethanol mixture drops into a nanofiber mat coating has been studied experimentally. Although the maximum imbibition area is affected by the deformation of the mats caused by the spreading of previous drops, this phenomenon is insignificant at the early stage of imbibition, which takes place in Lucas-Washburn regime. Comparison of the experimentally determined imbibition rate with the model prediction implies that the capillary rise in the corner between the substrate and the first layer of nanofibers is a possible mechanism contribution to imbibition. At the same time the results show that additional mechanisms should be taken into account to explain the observed behaviour, for example, the Marangoni effect.

\section{Acknowledgements}
\label{sec:thanks}


The authors thank the German Research Foundation (DFG) for financial support in the framework of Collaborative Research Center 1194 ``Interaction between Transport and Wetting Processes'', subproject A04.


\begin{thebibliography}{}


\bibitem{BARHATE.2007}
R. Barhate, S. Ramakrishna, Journal of Membrane Science  \textbf{296}, (2007) 1--8

\bibitem{Deng.2012}
X. Deng, L. Mammen, H.-J. Butt, D. Vollmer, Science  \textbf{335}, (2012) 67--70

\bibitem{Srikar.2009}
R. Srikar, T. Gambaryan-Roisman, C. Steffes, P. Stephan, C. Tropea, A. L. Yarin, International Journal of Heat and Mass Transfer \textbf{52}, (2014) 5814--5826

\bibitem{Alam2017imbibition}
E. Alam, S. Yadav, J. Schneider, T. Gambaryan-Roisman, Colloids Surf. A  \textbf{521}, (2017) 69--77

\bibitem{Bico.2001}
J. Bico, C. Tordeux, D. Qu{\'e}r{\'e}, Europhysics Letters  \textbf{55}, (2001) 214--220

\bibitem{Starov.2002}
V. M. Starov, S. R. Kostvintsev, V. D. Sobolev, M. G. Velarde, S. A. Zhdanov,  Journal of colloid and interface science \textbf{252}, (2002) 397--408

\bibitem{GambaryanRoisman.2014}
T. Gambaryan-Roisman, Current Opinion in Colloid {\&} Interface Science \textbf{19}, (2014) 320--335

\bibitem{Wemp2017water}
C. K. Wemp and V. P. Carey, Langmuir \textbf{33}, (2017) 14513--14525

\bibitem{Kumar2019spreading}
A. Kumar, J. Kleinen, J. Venzmer, A. Trybala, V. Starov, T. Gambaryan-Roisman, Colloids and Interfaces \textbf{3}, (2019) 53

\bibitem{Courbin.2009}
L. Courbin, J. C. Bird, M. Reyssat, H. A. Stone, Journal of physics. Condensed matter \textbf{21}, (2009) 464127

\bibitem{jun2013pool}
S. Jun, S. Sinha-Ray, A. Yarin, International Journal of Heat and Mass Transfer \textbf{62}, (2013) 99--11

\bibitem{Wang2016water}
Z. Wang, L. Espin, F. S. Bates, S. Kumar, C. W. Macoscko, Chemical Engineering Science \textbf{146}, (2016) 104--114

\bibitem{Freystein.2016}
M. Freystein, F. Kolberg, L. Spiegel, S. Sinha-Ray, R. P. Sahu, A. L. Yarin, T. Gambaryan-Roisman, P. Stephan, International Journal of Heat and Mass Transfer \textbf{93}, (2016) 827--833

\bibitem{SinhaRay2017pool}
 S. Sinha-Ray, W. Zhang, R.P. Sahu, S. Sinha-Ray, A. L. Yarin, International Journal of Heat and Mass Transfer \textbf{106}, (2017) 482--490
 
\bibitem{Fischer.2017}
S. Fischer, R. P. Sahu, S. Sinha-Ray, A. L. Yarin, T. Gambaryan-Roisman, P. Stephan, International Journal of Heat and Mass Transfer \textbf{108}, (2017) 2444--2450

\bibitem{Weickgenannt.2011}
C. M. Weickgenannt, Y. Zhang, A. Lembach, I. V. Roisman,  T. Gambaryan-Roisman, A. L. Yarin,  C. Tropea, Physical Review E \textbf{83}, (2011) 036305

\bibitem{weickgenannt2011inverse}
C. M. Weickgenannt, Y. Zhang, S. Sinha-Ray, I. V. Roisman,  T. Gambaryan-Roisman, C. Tropea, A. L. Yarin, Physical Review E \textbf{84}, (2011) 036310

\bibitem{Lembach.2010}
A. N. Lembach, H.-B-. Tan, I. V. Roisman, T. Gambaryan-Roisman, Y. Zhang,  C. Tropea,  A. L. Yarin,   Langmuir \textbf{26}, (2010) 9516--9523

\bibitem{Washburn.1921}
E. W. Washburn, Physical Review \textbf{17}, (1921) 273--283

\bibitem{Alleborn.2004}
N. Alleborn, H. Raszillier, J. Coll. Interface Sci.  \textbf{280}, (2004) 449--464

\bibitem{Romero.1996}
L. A. Romero, F. G. Yost, Journal of Fluid Mechanics \textbf{322}, (1996) 109--129

\bibitem{Rye1996flow}
R. R. Rye, J. A. Mann, F. G. Yost, Langmuir \textbf{12}, (1996), 555--565.

\bibitem{thammanna2018computations}
V. Thammanna Gurumurthy, D. Rettenmaier, I. V. Roisman, C. Tropea, S. Garoff, Colloids Surf. A \textbf{544}, (2018) 118--126

\bibitem{Mekhitarian.2017}
L. Mekhitarian, B. Sobac, S. Dehaeck, B. Haut, P. Colinet, Europhysics Letters  \textbf{120}, (2017) 16001

\bibitem{Kolliopoulos2019capillary}
P. Kolliopoulos, K. S. Jochem, R. K. Lade, Jr., L. F. Francis, S. Kumar, Langmuir \textbf{35}, (2019) 8131--8143

\bibitem{TGR2019imbibition}
T. Gambaryan-Roisman, Interfacial Phenom. Heat Transf. \textbf{7}, (2019) 239--253.


\bibitem{Vazquez1995surface}
G. V\'{a}zquez, E. Alvarez, J.M. Navaza, J. Chem. Eng. Data \textbf{40}, (1995) 611--614

\bibitem{Spencer2013contact}
S. J. Spencer, G. T. Andrews, C. G. Deacon, Semicond. Sci. Technol. \textbf{28}, (2013) 055011


\end{thebibliography}
\end{document}